\documentclass[preprint,11pt]{imsart}

\RequirePackage[numbers]{natbib}
\RequirePackage[colorlinks,citecolor=blue,urlcolor=blue]{hyperref}
\usepackage[utf8]{inputenc}
\usepackage[margin=3cm]{geometry}
\usepackage{amsmath,amssymb,amsthm,graphicx}
\usepackage{algorithm}
\usepackage{algpseudocode}
\usepackage{dsfont} 
\usepackage{booktabs}
\usepackage{caption}
\usepackage{subfigure}
\usepackage{subcaption}
\usepackage{tikz}
\usetikzlibrary{shapes,decorations,arrows,calc,arrows.meta,fit,positioning}
\tikzset{
          every node/.style={circle, inner sep=1.5mm, minimum size=0.55cm, draw, thick, black, fill=white, text=black},
          every path/.style={thick}
        }     
\usepackage{multirow} 
\usepackage{color}

\newtheorem{theorem}{Theorem}[section]

\newtheorem{example}[theorem]{Example}

\newcommand{\R}{\mathbb{R}}
\newcommand{\E}{\mathbb{E}}

\begin{document}
\begin{frontmatter}

\title{Identifying Total Causal Effects in Linear Models under Partial Homoscedasticity}
\runtitle{Identifying Total Causal Effects under Partial Homoscedasticity}

\author{\fnms{David} \snm{Strieder}\corref{}\ead[label=e1]{david.strieder@tum.de}}
\and
\author{\fnms{Mathias} \snm{Drton}\ead[label=e2]{mathias.drton@tum.de}}
\address{Munich Center for
Machine Learning and Department of Mathematics, \\ TUM School of Computation, Information and Technology, Technical University of Munich \\
	\printead{e1,e2}}

\runauthor{D. Strieder and M. Drton}

\begin{abstract}
A fundamental challenge of scientific research is inferring causal relations based on observed data. One commonly used approach involves utilizing structural causal models that postulate noisy functional relations among interacting variables. A directed graph naturally represents these models and reflects the underlying causal structure. However, classical identifiability results suggest that, without conducting additional experiments, this causal graph can only be identified up to a Markov equivalence class of indistinguishable models. Recent research has shown that focusing on linear relations with equal error variances can enable the identification of the causal structure from mere observational data. Nonetheless, practitioners are often primarily interested in the effects of specific interventions, rendering the complete identification of the causal structure unnecessary. In this work, we investigate the extent to which less restrictive assumptions of partial homoscedasticity are sufficient for identifying the causal effects of interest. Furthermore, we construct mathematically rigorous confidence regions for total causal effects under structure uncertainty and explore the performance gain of relying on stricter error assumptions in a simulation study.
\end{abstract}



\end{frontmatter}

\section{Introduction} 
\label{sec:intro}

Graphical models are a valuable tool for studying statistical dependencies among complex systems of random variables \citep{Lauritzen:book,Handbook}. In this paper, we study (directed) graphical models in their intuitive interpretation as structural causal models; edges indicate causal dependencies among interacting variables. Specifically, we consider statistical models specified by a recursive system of structural equations corresponding to a directed (acyclic) graph. The causal perspective views these structural equations as assignments rather than mathematical equations, reflecting the inherent asymmetry in cause-effect relationships.

A fundamental challenge of scientific research is inferring causal relations based on available data \citep{Spirtes:book,Pearl:book,Peters:book}. In particular, knowledge of the underlying causal structure is crucial to correctly estimate the effect of external interventions or to reason about counterfactual questions. However, expert knowledge of the exact causal mechanism governing the data-generating process is missing in many applied settings. Furthermore, performing classical controlled experiments to study the causal dependencies is often also not feasible due to ethical concerns or cost considerations. The field of causal discovery addresses this challenge by developing structure learning algorithms that estimate causal structures using only observed data \citep{Drton:17}.

A well-known result states that this causal structure, represented by a directed acyclic graph, can, at best, be identified up to a Markov equivalence class of indistinguishable models, unless one has access to interventional data or makes additional structural assumptions \cite[see, e.g.,][]{Drton:18}. Thus, an essential aspect of causal discovery is clarifying under which conditions the task is well-defined in that the causal quantities of interest can theoretically be identified from observational data alone. In this paper, we follow a line of research introduced by \citet{Peters:14} that focuses on linear relations and Gaussian errors with a common variance (see Section \ref{sec:back}). In this setting, the underlying causal mechanisms are fully identifiable from observational data alone, as the causal order is implied by ordering conditional variances \citep{Ghoshal:18, Chen:19}. However, practitioners are typically interested in specific causal effects, where identifying the entire causal structure seems excessive. This raises the question of whether less stringent modeling assumptions suffice to identify and estimate the effect of interest. Specifically, we investigate the identifiability of total causal effects in a setting that lies between the general case of unrestricted error variances and the fully homoscedastic case: We assume a shared variance only for the potential cause and the response of interest (Section \ref{sec:equal}). 

Identifiability aside, an important aspect of (causal) inference is appropriately accounting for uncertainty to draw reliable conclusions from causal estimates. With only access to observational data, this includes not only classical statistical uncertainty about the numerical size of involved effects but also causal structure uncertainty, stemming from the data-driven structure learning approach and the equivalence of multiple plausible models. In Section \ref{sec:inference}, we follow an ansatz first introduced in \citet{Strieder:21} and explicitly construct rigorous confidence regions for causal effects under partial homoscedasticity of error variances as well as for the general case. We conclude our work with a simulation study in Section \ref{sec:simulation} that compares the confidence regions under different error restrictions and thus exemplifies the performance gained by relying on stricter structural assumptions.

\section{Background} 
\label{sec:back}

In this section, we review structural causal models and the definition of the total causal effect, which is the causal target of interest in this study. Further, we recapitulate the notions of identifiability and Markov equivalence of models.

\subsection{Structural Causal Models and Total Causal Effects}

Structural Causal Models (SCMs) are a common tool to model noisy functional relations among a set of interacting variables $\{ X_i: i=1, \dots , d\}$. In an SCM, each variable $X_i$ is described as a function of a subset of other variables and a stochastic noise term $\varepsilon_i$. In this work, we focus on linear relations, that is, written in terms of random vectors, the data-generating process solves the linear equation system

\begin{equation}\label{eq:LSCM}
    \Vec{X}=B \Vec{X} + \Vec{\varepsilon}, 
\end{equation}
where $\Vec{X}=(X_1, \dots , X_d)$, $\Vec{\varepsilon}=(\varepsilon_1, \dots , \varepsilon_d)$, and $B:=[\beta_{j,i}]_{j,i=1}^d$ is a matrix that represents direct causal dependencies. We further assume that the errors are normal distributed, that is, $\varepsilon\sim\mathcal{N}(\Vec{0},\Omega)$, where $\Omega=\text{diag}(\Vec{\omega})$ with $\Vec{\omega}=(\omega_1, \dots , \omega_d)$. Note that the errors are assumed to be uncorrelated, which implies causal sufficiency. The causal perspective arises when viewing these equation equations as making assignments rather than mathematical equations. Thus, the underlying causal structure of SCMs is naturally represented by a (minimal) directed graph $G$, where each node $i$ in the graph corresponds to a variable $X_i$ and edges $i \rightarrow j$ indicate direct causal dependencies, that is, $\beta_{j,i}\neq 0$. Throughout this work, we focus on acyclic causal relations, which entails that $B$ is permutation similar to a lower triangular matrix and yields the unique solution $\Vec{X}=(I_d-B)^{-1} \Vec{\varepsilon}$ of the equation system \eqref{eq:LSCM}, where $I_d$ denotes the $d\times d$ identity matrix. Thus, $\Vec{X} \sim \mathcal{N}(\Vec{0},\Sigma)$ where the covariance matrix is given by 
\[
\Sigma=(I_d-B)^{-1}\Omega(I_d-B)^{-T}.
\]

We use the following notation and graphical concepts in the remainder of the article. We write $i<_{G}j$ if node $i$ precedes node $j$ in a causal ordering of the corresponding directed acyclic graph (DAG). If the DAG contains an edge from node $i$ to node $j$, then node $i$ is called a \textit{parent} of node $j$, and we denote the set of all parents of node $j$ with $p(j)$. Further, if the DAG contains a directed path from node $i$ to node $j$, then node $j$ is called a \textit{descendant} of node $i$, and we denote the set of all descendants of node $i$ with $d(i)$. Finally, we write $\Sigma_{j,i|p(i)}$ for the conditional covariance matrix, that is,
    \begin{equation*}
        \Sigma_{j,i|p(i)}:=\Sigma_{j,i}-\Sigma_{j,p(i)}(\Sigma_{p(i),p(i)})^{-1}\Sigma_{p(i),i}.
    \end{equation*}

The target of interest in this work is the \textit{total causal effect} $\mathcal{C}(i\rightarrow j)$, that is, the effect of an external intervention on $X_i$ onto $X_j$. Within the framework of linear SCMs, the total causal effect is formally defined as the unit change in the expectation of $X_j$ with respect to an intervention in $X_i$
\begin{equation*} 
   \mathcal{C}(i \rightarrow j):=\frac{\text{d}}{\text{d} x_i} \E[X_j|\text{ do}(X_i=x_i)].
\end{equation*}
If the underlying DAG $G$ is known, this parameter of interest can be expressed as a simple function of the covariance matrix $\varphi(G,\Sigma):=\Sigma_{j,i|p(i)}(\Sigma_{i,i|p(i)})^{-1}$, which is the regression coefficient of $X_i$ when regressing
$X_j$ on $(X_i, X_{p(i)})$. However, difficulties arise in practice when the underlying causal structure is unknown, and thus, a valid adjustment set is unknown. 

\subsection{Markov Equivalence and Identifiability}

An important question is, thus, given an observational distribution $P_{\Sigma}$ from a set of possible distributions $\{P_{\Sigma}: \Sigma \in \mathcal{M}\}$, can one uniquely recover the (causal) target of interest $\varphi(G,\Sigma)$? Each DAG $G$ may generate a (sub-)set of possible distributions from the model space $\mathcal{M}$ with entailing effect $\varphi(G,\Sigma)$, and thus, geometrically, this question is about the set of distributions that lie in the intersection of multiple plausible (causal) models. 

Generally, an underlying linear Gaussian SCM naturally encodes a set of conditional independence constraints, given graphically by the set of d-separation relations in the underlying (true) DAG, which correspond algebraically to polynomial constraints on the set of possible covariance matrices. Two DAGs are called \textit{Markov equivalent} if they imply the same set of conditional independence constraints and thus generate the same set of distributions. However, these Markov equivalent DAGs do not necessarily entail the same causal effect of interest and, therefore, without additional (structural) assumptions, identification is at most possible up to a Markov equivalence class, that is, a set of potential values of the target of interest, each corresponding to a DAG within the same Markov equivalence class.  

Additional faithfulness-type assumptions commonly guarantee at least being able to identify the Markov equivalence class by assuming that the given distribution does not satisfy additional conditional independence constraints, which are not implied by d-separation relations in the (true) DAG. We note that to recover our target of interest, the total causal effect, it is not necessary to fully identify the underlying DAG. We do not need to be able to distinguish a generating (minimal) DAG from its supergraphs to identify the target effect since all supergraphs yield a valid adjustment set for the total causal effect. To see that, note that every 'additional' parent of node $i$ in a supergraph has to be conditionally independent of node $i$ given the (minimal) parent set, which follows from d-separation in the minimal DAG. Thus, while assuming sparsity might improve computational aspects (by reducing the size of the model space), from a theoretical point of view, it suffices to only consider complete DAGs, that is, (causal) topological orderings of the involved variables, which includes all (sparse) DAGs as subgraphs.

In this article, we will focus our attention on the concept of \textit{generic identifiability}. We call the causal parameter $\mathcal{C}(i \rightarrow j)$ generically identifiable if it can be recovered uniquely for almost all $\Sigma \in \mathcal{M}$ from a given observational distribution $P_{\Sigma}$ out of a set of possible distributions $\{P_{\Sigma}: \Sigma \in \mathcal{M}\}$. Put differently, the set of covariance matrices corresponding to distributions for which the target of interest is not identifiable form a Lebesgue measure zero subset of our model space $\mathcal{M}$.

\section{Homoscedasticity in SCMs} 
\label{sec:equal}

In this section, we investigate how introducing homoscedasticity constraints among the different error variances leads to identifiability of the (causal) parameters of interest. As mentioned before, in general, identification is at most possible up to a set of indistinguishable effects corresponding to Markov equivalent models. To be precise, in our setting under the assumption of an underlying linear Gaussian SCM, every possible distribution $\mathcal{N}(0,\Sigma)$ belongs to a covariance matrix from the set $\mathcal{M}:=\bigcup_{G \in \mathcal{G}(d)} \mathcal{M}(G)$, where $\mathcal{G}(d)$ is the set of all complete DAGs on $d$ nodes and
\begin{equation}\label{eq:parameterized}
    \mathcal{M}(G)=\Big\{\Sigma \in \text{PD}(d): \exists B \in \R^{G} , \Vec{\omega} \in \R^d \text{ with } 
	   \Sigma=(I_d-B)^{-1}\text{diag}(\Vec{\omega})(I_d-B)^{-T} \Big\}
\end{equation}
is the (causal) model given by a DAG $G$, where $\R^G=\{B \in \R^{d \times d} : \beta_{j,i}=0 \text{ if } j <_G i \}$. Note that without restrictions on the error variances, each set $\mathcal{M}(G)$ corresponds to the entire cone of positive definite matrices $\text{PD}(d)$. In other words, all complete DAGs are Markov equivalent, and any complete DAG could have generated every possible given observational distribution. However, the total causal effect given by $\varphi(G,\Sigma)$ is not the same within all complete DAGs $G$ that could have generated $\Sigma$. Thus, the total causal effect is not identifiable without any additional order or structure constraints.

\begin{theorem}[see, e.g., \cite{Pearl:book}]
    The total causal effect $\mathcal{C}(i \rightarrow j)$ is not generically identifiable under the assumption of an underlying linear Gaussian SCM.
\end{theorem}

Nevertheless, given an observational distribution $P_{\Sigma}$, we can recover a finite set of possible values of the total causal effects by enumerating all effects corresponding to all equivalent complete DAGs \citep{Maathuis:09}. We emphasize that while the number of permutations of $d$ nodes is $d$ factorial, many of these permutations lead to the same effects. For example, half of the permutations correspond to orderings with $j <_G i$ and, thus, entail a zero-sized effect. Similarly, the ordering among the parent set $p(i)$ or among the descendant set $d(i)$ is irrelevant to the specific value of the total causal effect. In fact, the size of the set of possible values of the total causal effect is upper bound by $2^{(d-2)}+1$. However, under a general linear Gaussian SCM, this set of indistinguishable effects contains zero and non-zero values and, thus, is not informative about the existence or direction of a total causal effect without additional assumptions.

\subsection{Equal Error Variance Constraint}

A frequently used structural assumption that resolves this issue is introducing homoscedasticity among the error variances, that is, assuming $\omega_i = \omega_j$ for all $i,j= 1, \dots ,d$. This assumption restricts the set of possible distributions to a proper subset of the cone of positive definite matrices. Working with this assumption of equal error variances does not seem unreasonable for exploratory analyses in applications involving variables from similar domains or measurements obtained by similar machines and applications with concrete knowledge about external error variances. 

From a causal perspective, this restriction refines the Markov equivalence classes and yields a straightforward criterion to identify the underlying causal ordering. Since the external error variances are equal, source nodes in the DAG have to be of minimal variance. To be precise, the homoscedasticity assumption corresponds to the algebraic constraint that the variance of each node conditioned on its parents has to be equal. Thus, for a given (complete) DAG, the set of possible distributions is given by 
\begin{equation*}
    \mathcal{M}_{EV}(G)=\Big\{\Sigma \in \text{PD}(d): \exists  \omega >0 \text{ with } 
	    \omega=\Sigma_{k,k|p(k)} \quad  \forall\  k=1, \dots , d  \Big\}.
\end{equation*} 
Note that any given $\Sigma$ has a minimal diagonal element, which 'fixes' the value of the equal error variance $\omega$ and has to correspond to a source node in an underlying (minimal) DAG. If this minimal element is not unique, all nodes corresponding to minimal elements must be source nodes. By recursively conditioning on previous nodes and selecting nodes with minimal conditional variances, we obtain a unique one-to-one correspondence between a distribution $\Sigma$ and a generating underlying minimal DAG. Thus, two causal models $\mathcal{M}_{EV}(G)$ only coincide at distributions generated by subgraphs that admit both complete graphs as valid causal orderings. However, since the underlying minimal DAG is unique, all valid causal orderings entail the same total causal effect. As mentioned, this follows due to d-separation criteria in the (unique) minimal DAG. Every 'additional' parent of node $i$ in a supergraph has to be conditionally independent of node $i$ given the (minimal) parent set. In other words, we obtain the following result.

\begin{theorem}[\cite{Peters:14}]
    The total causal effect $\mathcal{C}(i \rightarrow j)$ is globally identifiable under the assumption of an underlying linear Gaussian SCM with full homoscedasticity.
\end{theorem}

We emphasize that this holds globally without additional faithfulness-type assumptions.

\subsection{Identifiability under Partial Homoscedasticity}

In this section, we answer the question of what happens between the general case of arbitrary error variances and the fully homoscedastic case with one common error variance. Specifically, we study the case of partial homoscedasticity; that is, we restrict $\omega_i=\omega_j$ only for potential cause $X_i$ and response $X_j$ of interest. This extends the possible applications of the framework towards cases where (only) the potential cause and response are from similar domains and cases with concrete knowledge about external error variances of (only) the potential cause and response.
Algebraically, this assumption of partial homoscedasticity corresponds to the constraint that the variance of node $i$ conditioned on its parents equals the variance of node $j$ conditioned on its parents. Thus, for a given (complete) DAG, the causal model under partial homoscedasticity is given by

\begin{equation*}
    \mathcal{M}_{PEV}(G)=\Big\{\Sigma \in \text{PD}(d): \Sigma_{i,i|p(i)}=\Sigma_{j,j|p(j)}  \Big\}.
\end{equation*} 
In previous work, \citet{Wu:23} show that this assumption refines the Markov equivalence classes in such a way that two equivalent DAGs only generate the same causal model if and only if both postulate the same parent sets $p(i)$ and $p(j)$, respectively. This raises the question of whether this structural partial homoscedasticity assumption is enough to uniquely recover the total causal effect of interest $\mathcal{C}(i \rightarrow j)$ from a given observational distribution $P_{\Sigma}$. To answer that question, we study the set of possible values of the total causal effect entailed by a distribution that lies in the intersection of multiple plausible causal models.

Without loss of generality, we focus on the intersection between two (complete) graphs $G^1$ and $G^2$, that is, the intersection between two models $\mathcal{M}_{PEV}(G^1)$ and $\mathcal{M}_{PEV}(G^2)$. Both models can be parametrized similarly to \eqref{eq:parameterized}, with $B \in \R^G$ and $\Vec{\omega} \in \R^{d-1}$ under the partial homoscedasticity restriction $\omega_i=\omega_j$. Thus, as parameterized models, $\mathcal{M}_{PEV}(G^1)$ and $\mathcal{M}_{PEV}(G^2)$ are irreducible algebraic models \cite[see, e.g.,][]{Cox:15}. The intersection of two irreducible models is either of lower dimension or both models fully coincide. As already mentioned, Theorem 4.1 of \cite{Wu:23} states that the two models fully coincide if and only if they imply the same parent sets for $i$ and $j$, respectively, that is, $p_{G^1}(i)=p_{G^2}(i)$ and $p_{G^1}(j)=p_{G^2}(j)$. In this case however, both models would also entail the same total causal effect $\varphi(G^1,\Sigma)=\varphi(G^2,\Sigma)=\Sigma_{j,i|p(i)}(\Sigma_{i,i|p(i)})^{-1}$ since $p_{G^1}(i)=p_{G^2}(i)$. In the other case, if $G^1$ and $G^2$ do not imply the same parent sets for $i$ and $j$, the intersection of $\mathcal{M}_{PEV}(G^1)$ and $\mathcal{M}_{PEV}(G^2)$ consequently has to be lower dimensional and, thus, a Lebesgue measure zero subset. Nevertheless, this intersection is not empty; see Example \ref{example}. Combining both cases, we obtain the following result.

\begin{theorem}
    The total causal effect $\mathcal{C}(i \rightarrow j)$ is generically identifiable under the assumption of an underlying linear Gaussian SCM with partial homoscedasticity among $i$ and $j$.
\end{theorem}

In other words, we can uniquely recover the total causal effect of interest for almost all $\Sigma \in \bigcup_{G \in \mathcal{G}(d)} \mathcal{M}_{PEV}(G)$, that is, for almost all observational distributions from a linear Gaussian SCM under partial homoscedasticity. See Appendix \ref{apd:example2} for an explicit example. Further, note that analogous additional faithfulness-type assumptions, i.e., assuming that the distribution satisfies only the equal error variance constraint between $i$ and $j$ as implied by the parent set of the true underlying DAG, would guarantee global identifiability. However, we emphasize that we can explicitly construct not only 'unfaithful' distributions in the intersection between two models that would imply effects of different magnitudes but also distributions in the intersection of two models that imply different directions for the considered effect. Thus, not even the causal ordering of the potential cause $X_i$ and response $X_j$ is globally identifiable without additional assumptions.
 
\begin{example}\label{example}
The two linear Gaussian SCMs with partial homoscedasticity $\omega_1=\omega_2$, given by the underlying DAGs $G^1$ and $G^2$ generate the same observational distribution (with $\Vec{\omega}=(1,1,1)$ and $\Vec{\omega}=(0.81,0.81,1.52)$, respectively) but entail a different total causal effect $\mathcal{C}(1\rightarrow 2)$, that is, $\varphi(G^1,\Sigma)=0.29\neq 0=\varphi(G^2,\Sigma)$. 
\begin{figure}[ht]
\centering
\subfigure[$G^1$]{
        \begin{tikzpicture}[->,>=triangle 45,shorten >=1pt,
          auto,thick, main
          node/.style={circle,fill=gray!10,draw}]
            \node[main node,draw=blue] (1) at (0,0)     {$1$};
            \node[main node,draw=blue] (2) at (2,1)    {$2$};
            \node[main node] (3) at (2,-1)     {$3$};
            \path[color=red,every node/.style={font=\sffamily\small}] 
            (1) edge node[xshift=0.5mm] {$0.29$} (2);
            \path[every node/.style={font=\sffamily\small}]
            (1) edge node[below,xshift=-5mm] {$-0.38$} (3);
            \path[every node/.style={font=\sffamily\small}]
            (2) edge node[yshift=0.5mm] {$0.70$} (3);
        \end{tikzpicture}
}
    \hspace{1cm}
\subfigure[$G^2$]{
        \begin{tikzpicture}[->,>=triangle 45,shorten >=1pt,
          auto,thick, main
          node/.style={circle,fill=gray!10,draw}]
            \node[main node,draw=blue] (1) at (0,0)     {$1$};
            \node[main node,draw=blue] (2) at (2,1)    {$2$};
            \node[main node] (3) at (2,-1)     {$3$};
            \path[every node/.style={font=\sffamily\small}] 
            (3) edge node[right,yshift=-0.5mm] {$0.43$} (2);
            \path[color=red,every node/.style={font=\sffamily\small}] 
            (2) edge node[above,xshift=-1mm,yshift=1mm] {$0.46$} (1);
            \path[every node/.style={font=\sffamily\small}]
            (3) edge node[yshift=-0.5mm,xshift=2mm] {$-0.31$} (1);
        \end{tikzpicture}
    }
\end{figure}

We used computer algebra software to find a covariance matrix that satisfies both equal error variance constraints $\Sigma_{1,1}=\Sigma_{2,2|1}$ and $\Sigma_{1,1|\{2,3\}}=\Sigma_{2,2|3}$. Thus, such a covariance could have been generated by both causal orderings $1 <_G 2 <_G 3$ and $3 <_G 2 <_G 1$ and exemplifies that not even the direction of the effect is globally identifiable. Note that we rounded the displayed values; for more information see Appendix \ref{apd:example}.
\end{example}

\section{Causal Inference under Partial Homoscedasticity} 
\label{sec:inference}
Theoretical identifiability aside, intuitively, stronger structural assumptions should lead to more informative inference results. Thus, in the following sections, we want to investigate how causal inference differs under these three error variance assumptions and how much performance can be gained by relying on stricter structural assumptions. 
 
\subsection{Maximum Dual Likelihood Estimation}\label{sec:estimation}

Assume we have access to observational data in form of $n$ independent copies  $\Vec{X}^{(1)},\dots, \Vec{X}^{(n)}$ of a random vector $\Vec{X}$ that follows a linear Gaussian SCM, given by an unknown DAG $G$. This section aims to estimate the set of indistinguishable total causal effects $\mathcal{C}(i \rightarrow j)$, corresponding to all equivalent models, similar to the IDA-framework \citep{Maathuis:09}. We employ dual likelihood methods \citep{Brown:86,Kauermann:96} as an alternative to classical maximum likelihood estimation due to computational advantages \citep{Strieder:24}. The main idea of dual likelihood is to swap the arguments in the usual classical minimization of the Kullback–Leibler divergence, which yields a different estimator and corresponds to maximizing 
\begin{equation*}
   \ell^{dual}_n(\Sigma):=-\log \det (\Sigma^{-1}) - \mathrm{tr}(\Sigma\widehat{\Sigma}^{-1}),
\end{equation*}
where $\widehat{\Sigma} = \frac{1}{n} \sum_{l=1}^n \Vec{X^{(l)}} (\Vec{X^{(l)}})^T$ is the empirical covariance. We note that employing dual likelihood transfers the optimization problem from a covariance matrix to its inverse, simplifying constraints on total causal effects to facilitate the test inversion in Section \ref{sec:intervals}. 

Similar to the classical likelihood, $\widehat{\Sigma}$ is the unconstrained dual maximum likelihood estimate for $\Sigma \in \mathcal{M}$. Thus, in the general case without any restrictions on the error variances, the total causal effect can be estimated via $\{\varphi(G,\widehat{\Sigma}) \in \R : G \in \mathcal{G}(d) \},$ which is the set of all causal effects implied by all indistinguishable models that could have generated the maximum (dual) likelihood estimate $\widehat{\Sigma}$. Note that many complete DAGs necessarily entail the same effect as they imply the same adjustment set $p(i)$. Thus, to calculate this proposed estimate, we only need to compute $2^{d-2}$ values, that is, enumerate all distinct parent sets $p(i)$ that do not include $j$, and further append zero. Furthermore, note that the adjustment sets implied by complete DAGs are not necessarily efficient if the true underlying DAG is not complete \citep{Henckel:22}.

Under the additional structural assumption of partial homoscedasticity between potential cause $i$ and response $j$, that is, assuming $\omega_i=\omega_j$, the maximum likelihood estimation is a complex combinatorial optimization problem. However, for a fixed DAG $G$, dual maximum likelihood estimation is equivalent to solving a sequence of linear regression problems for each node regressed on its descendants. Thus, we obtain the optimal value
\begin{equation*}
    \sup_{\Sigma \in \mathcal{M}_{PEV}(G)}\ell^{dual}_n(\Sigma)=-\sum_{k \neq i,j}^d\log\big((\widehat{\Sigma}^{-1})_{k,k|d(k)}\big)-2\log\big(\tfrac{1}{2}\big((\widehat{\Sigma}^{-1})_{i,i|d(i)}+(\widehat{\Sigma}^{-1})_{j,j|d(j)}\big)\big) -d.
\end{equation*}
The corresponding total causal effect is the regression coefficient of node $j$ when regressing node $i$ on its descendants, that is, $\tau_j\big(\big((\widehat{\Sigma}^{-1})_{d(i),d(i)}\big)^{-1}(\widehat{\Sigma}^{-1})_{d(i),i}\big),$ where $\tau_j$ projects the $|d(i)|$-dimensional vector onto the component corresponding to $j$ if $j\in d(i)$ and zero otherwise. Subsequently, maximizing over the space of complete DAGs $G \in \mathcal{G}$ yields the dual maximum likelihood estimate under partial homoscedasticity. We emphasize that this dual maximum likelihood estimate may lie in the intersection of multiple causal models implying different total causal effects. Thus, analogously to the general case, we propose to estimate the total causal effect under structure uncertainty as the set of all causal effects corresponding to DAGs that achieve this optimal value. Using combinatorial shortcuts, we can compute this proposed estimate without checking all $d$ factorial permutations, i.e., all complete DAGs implying the same parent sets $p(i)$ and $p(j)$ are equivalent and entail the same causal effect of interest. Thus, it suffices to compute and compare two times $3^{d-2}$ optimal values.

Further, under full homoscedasticity among all error variances and for a fixed DAG $G$, we obtain the optimal value 
\begin{equation*}
    \sup_{\Sigma \in \mathcal{M}_{EV}(G)}\ell^{dual}_n(\Sigma)=-d\log\big(\tfrac{1}{d}\sum_{k=1}^d (\widehat{\Sigma}^{-1})_{k,k|d(k)}\big) -d
\end{equation*}
with corresponding total causal effect $\tau_j\big(\big((\widehat{\Sigma}^{-1})_{d(i),d(i)}\big)^{-1}(\widehat{\Sigma}^{-1})_{d(i),i}\big)$. Analogously, maximizing over the space of complete DAGs $G \in \mathcal{G}$ then yields the dual maximum likelihood estimate. This dual maximum likelihood estimate may also lie in the intersection of multiple causal models. However, in this case, this intersection has to correspond to distributions generated by a lower-dimensional subgraph. Thus, all complete DAGs that achieve this optimal value are valid causal orderings of the true underlying subgraph and entail the same total causal effect. We note that leveraging combinatorial shortcuts to calculate this total causal effect estimate is more complex, but evaluating all permutations can be avoided with ideas similar to \citet{Silander:06}.

\subsection{Confidence Intervals for Total Causal Effects}\label{sec:intervals}

In order to draw reliable conclusions from causal estimates under structure uncertainty, we further want to construct a confidence interval, that is, in the technical sense, a region that (at least asymptotically) guarantees a desired frequentist coverage probability of the true total causal effect of interest. Recent proposals that rigorously account for uncertainty in causal inference rely on test-inversion approaches to construct confidence sets for causal quantities of interest \citep{Strieder:23,Wang:23}. \citet{Strieder:23} explicitly construct a confidence region for the total causal effect by inverting joint tests for the underlying causal structure and effect size in the concrete setting of an underlying linear Gaussian SCM with equal error variances. We will take up their methodology and extend it to the partial homoscedastic and general cases. Again, we emphasize that the total causal effect is not globally identifiable when departing from the complete homoscedastic case. Thus, structure uncertainty is not only aleatoric due to the data-driven model choice but also epistemic due to multiple models being indistinguishable. We aim to construct reliable confidence intervals that consider all sources of uncertainty and cover all indistinguishable effects with the desired frequency.

In the general setting, without assuming equal error variance, the ansatz of inverting joint tests for causal structure and effect size leads to the following statistical testing problem for all $\psi \in \R$
\begin{equation*}
    H_0^{(\psi)} :
	    \Sigma \in \mathcal{M}^{\psi} \quad \text{against} \quad H_1 : \Sigma \in \mathcal{M} \backslash \mathcal{M}^{\psi},
\end{equation*}
where $\mathcal{M}^{\psi}:=\bigcup_{G \in \mathcal{G}(d)} \mathcal{M}^{\psi}(G)$ with $\mathcal{M}^{\psi}(G):=\{\Sigma \in \text{PD}(d): \varphi(G,\Sigma)=\psi \}$. We solve the testing problem using the dual likelihood ratio test and the theory of intersection union tests \cite[see, e.g.,][]{casella:book} to obtain the following result.

\begin{theorem}\label{thm:confgen}
 Let $\alpha \in (0,1)$. For all $G \in \mathcal{G}(d)$ with $i <_G j$ define
 \begin{equation*}
        D(G):= (\widehat{\Sigma}^{-1})_{j,i|d(i)\setminus \{j\}}^2-(\widehat{\Sigma}^{-1})_{j,j|d(i)\setminus \{j\}}\Big((\widehat{\Sigma}^{-1})_{i,i|d(i)\setminus \{j\}} - (\widehat{\Sigma}^{-1})_{i,i|d(i)} \exp\big(\tfrac{\chi^2_{1,1-\alpha}}{n}\big) \Big).
    \end{equation*}
    Then an asymptotic $(1-\alpha)$-confidence set for the total causal effect $\mathcal{C}(i\rightarrow j)$ under the assumption of an underlying linear Gaussian SCM is given by
    \begin{equation*}
            C :=\bigcup_{G \in \mathcal{G}(d)\, :\, D(G)\geq0} \big[L(G), U(G)\big]  \bigcup \big\{0 \big\}, 
    \end{equation*}    
     where the closed-form solution for the limits is given by 
    \begin{equation*}
        L(G):=\frac{-(\widehat{\Sigma}^{-1})_{j,i|d(i)\setminus \{j\}}  - \sqrt{D(G)}}{(\widehat{\Sigma}^{-1})_{j,j|d(i)\setminus \{j\}}},  \qquad U(G):=\frac{-(\widehat{\Sigma}^{-1})_{i,j|d(i)\setminus \{j\}}  + \sqrt{D(G)}}{(\widehat{\Sigma}^{-1})_{j,j|d(i)\setminus \{j\}}}.
    \end{equation*}
\end{theorem}
The proof is given in Appendix \ref{apd:proofconfgen}.

Without any additional structural assumption on the error variances, the total causal effect is not identifiable, and all complete DAGs could have generated every possible observational distribution. While the confidence regions still contain information and restrict the real line of all possible effects, they will always include the possibility of no effect. 

Inducing additional structure by assuming equal error variances among the potential cause $i$ and response $j$ (potentially) resolves this ambiguity and leads to confidence regions via inverting the following statistical testing problem for all $\psi \in \R$
\begin{equation*}
    H_0^{(\psi)} :
	    \Sigma \in \mathcal{M}_{PEV}^{\psi} \quad \text{against} \quad H_1 : \Sigma \in \mathcal{M}_{PEV} \backslash \mathcal{M}_{PEV}^{\psi},
\end{equation*}
where $\mathcal{M}_{PEV}^{\psi}:=\bigcup_{G \in \mathcal{G}(d)} \mathcal{M}_{PEV}^{\psi}(G)$ with $\mathcal{M}_{PEV}^{\psi}(G):=\{\Sigma \in  \mathcal{M}_{PEV}(G): \varphi(G,\Sigma)=\psi \}$. Analogously, solving this testing problem with dual likelihood ratio tests and the theory of intersection union tests yields the following result.

\begin{theorem}\label{thm:confpart}
 Let $\alpha \in (0,1)$. Define $K:=\min_{G \in \mathcal{G}(d)} K(G)$ and $Z:=\min_{G \in \mathcal{G}(d): j <_G i} K(G)$, where
\begin{equation*}
        K(G):= \prod^d_{k\neq i,j}\sqrt{(\widehat{\Sigma}^{-1})_{k,k|d(k)} } \Big((\widehat{\Sigma}^{-1})_{i,i|d(i)}+(\widehat{\Sigma}^{-1})_{j,j|d(j)}\Big).
    \end{equation*}
 Furthermore, for all $G \in \mathcal{G}(d)$ with $i <_G j$ define
\begin{align*}
        D(G):= &(\widehat{\Sigma}^{-1})_{j,i|d(i)\setminus \{j\}}^2\\ &-(\widehat{\Sigma}^{-1})_{j,j|d(i)\setminus \{j\}}\Bigg((\widehat{\Sigma}^{-1})_{j,j|d(j)}+(\widehat{\Sigma}^{-1})_{i,i|d(i)\setminus \{j\}} -\frac{K \exp\big(\tfrac{\chi^2_{2,1-\alpha}}{2n}\big)}{\prod^d_{k\neq i,j}\sqrt{(\widehat{\Sigma}^{-1})_{k,k|d(k)} }} \Bigg).
\end{align*}    
 Then an asymptotic $(1-\alpha)$-confidence set for the total causal effect $\mathcal{C}(i\rightarrow j)$ under the assumption of an underlying linear Gaussian SCM with (partially) equal error variance among $i$ and $j$ is given by
    \begin{equation*}
            C :=\bigcup_{G \in \mathcal{G}(d)\, :\, D(G)\geq 0} \big[L(G), U(G)\big]  \bigcup \big\{0: Z \leq K \exp\big(\tfrac{\chi^2_{1,1-\alpha}}{2n}\big) \big\}, 
    \end{equation*}    
    where the closed-form solution for the limits is given by 
    \begin{equation*}
        L(G):=\frac{-(\widehat{\Sigma}^{-1})_{j,i|d(i)\setminus \{j\}}  - \sqrt{D(G)}}{(\widehat{\Sigma}^{-1})_{j,j|d(i)\setminus \{j\}}},  \qquad U(G):=\frac{-(\widehat{\Sigma}^{-1})_{i,j|d(i)\setminus \{j\}}  + \sqrt{D(G)}}{(\widehat{\Sigma}^{-1})_{j,j|d(i)\setminus \{j\}}}.
    \end{equation*}
\end{theorem}
The proof is given in Appendix \ref{apd:proofconfpart}.

We highlight the overarching structure of the confidence regions, which consist of intervals that each represent a plausible causal ordering with $ i <_G j$ and possibly a zero that reflects remaining uncertainty about the direction of the effect. It is worth noting that any causal ordering with $ i <_G j$ immediately implies no effect. Therefore, informally, the hypothesis of no effect for a DAG with $ i <_G j$ is larger than that of a specific effect $\psi$ in a DAG with $ j <_G i$. This intuitively explains the lower degrees of freedom in the corresponding critical value.

\section{Simulation Study} 
\label{sec:simulation}

\begin{figure}[t]
    \centering
    \includegraphics[width=0.9\linewidth]{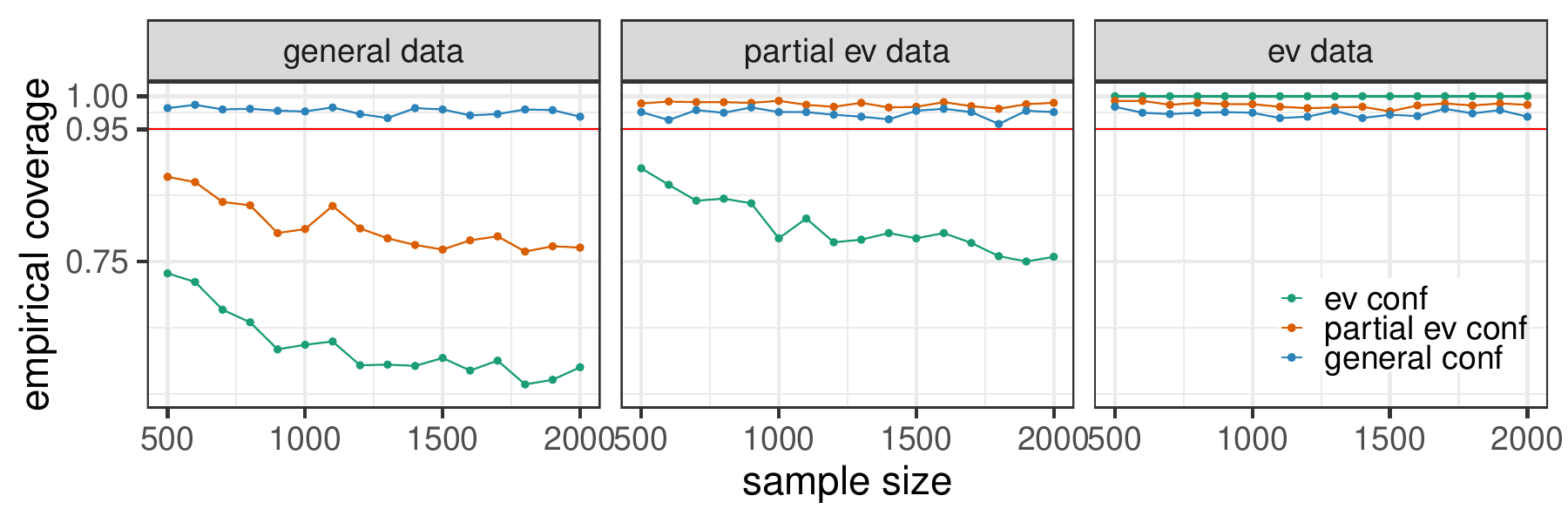}
    \caption{Empirical coverage of $95\%$-confidence intervals for the total causal effect.}\label{fig:cover10}
\end{figure}
\begin{figure}[t]
    \centering
    \includegraphics[width=0.9\linewidth]{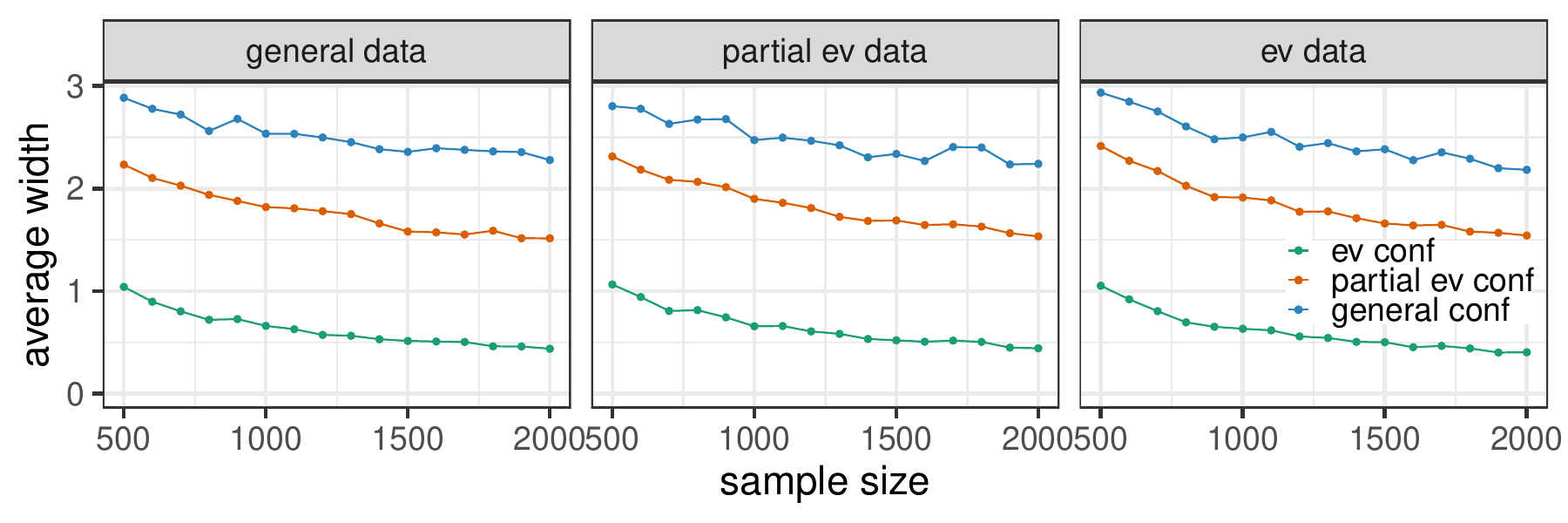}
    \caption{Mean width of $95\%$-confidence intervals for the total causal effect.}\label{fig:width10}
\end{figure}
\begin{figure}[t]
    \centering
    \includegraphics[width=0.9\linewidth]{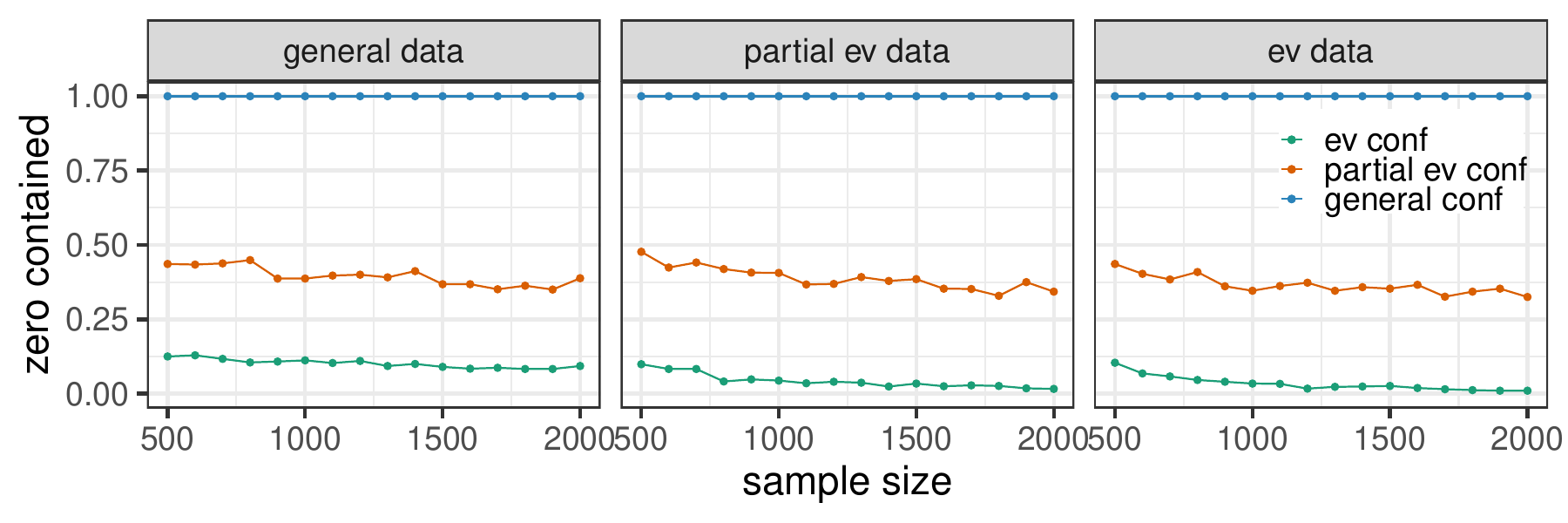}
    \caption{Proportion of times zero contained in $95\%$-confidence intervals for the total causal effect.}\label{fig:zero10}
\end{figure}

In this section, we present the results of a simulation study that compares the performance of our proposed confidence regions (\texttt{general conf} with no variance assumption and \texttt{partial ev conf} assuming partial homoscedasticity) and the confidence region proposed in \citet{Strieder:24} under the (strict) assumption of all error variances being equal (\texttt{ev conf}). Thus, our simulations contrast the potential information gain of relying on stricter structural error variance assumptions. Our experiments were designed following \citet{Strieder:24} with synthetic data from $10-$dimensional sparse DAGs with true non-zero total causal effects under the three error variance regimes.  For details on the data generation process and additional simulation results for data with no true effect, see Appendix \ref{apd:sim}. 

 Figure \ref{fig:cover10} reports the empirical coverage frequencies. As expected, all methods achieve the desired coverage frequency of $0.95$ under their respective error variance regime (and more restrictive cases); however, they fail under the more general regimes. To investigate the potential upside of relying on stricter variance assumptions, we investigate how conclusive the proposed confidence regions are about the existence and size of causal effects in the form of the average width of the non-zero part of the confidence sets (Figure \ref{fig:width10}) as well as proportions of times zero is contained in the confidence sets (Figure \ref{fig:zero10}). Note that the methods do not adapt to the complexity of the data; that is, each method's width and zero proportions do not vary across the different data generation settings. Furthermore, without any error variance assumption, the confidence region is never conclusive about the existence of an effect, highlighting that structural assumptions are necessary to obtain decisive results. Nevertheless, the validity of the confidence statements relies on the correctness of the error variance assumption, which thus, must be carefully chosen to make calibrated and informative confidence statements about total causal effects under structure uncertainty.

\section{Conclusion} 
\label{sec:discussion}

An essential aspect of causal discovery is clarifying under which conditions it is theoretically possible to identify causal quantities from observational data alone. In this paper, we investigate how different assumptions on the error variances lead to the (non-)identifiability of total causal effects of interest. While the classical setting with arbitrary error variances leads to a (finite) set of indistinguishable potential effects,  the recently studied equal error variance setup with one common variance entails global identifiability. We demonstrate that in between those cases,  assuming partial equal variance between the potential cause and response of interest is sufficient to generically identify the total causal effect. 

Furthermore, appropriately accounting for uncertainty is crucial to draw reliable conclusions from causal estimates. With only access to observational data, this includes classical statistical uncertainty about the numerical size of involved effects and causal structure uncertainty, stemming from a data-driven structure learning approach as well as the equivalence of multiple plausible models. Via a test inversion approach, we construct rigorous confidence regions that account for all types of uncertainty and show that if practitioners are willing to rely on stricter error variance assumptions, then the corresponding confidence regions are more informative about the existence and size of total causal effects.

\begin{acks}[Acknowledgments]
This project has received funding from the European Research Council (ERC) under the European Union’s Horizon 2020 research and innovation programme (grant agreement No. 83818). Further, this work has been funded by the German Federal Ministry of Education and Research and the Bavarian State Ministry for Science and the Arts. The authors of this work take full responsibility for its content.
\end{acks}

\newpage

\bibliographystyle{imsart-nameyear}
\bibliography{bibliography.bib}

\begin{thebibliography}{23}

\bibitem[\protect\citeauthoryear{Brown}{1986}]{Brown:86}
\begin{bbook}[author]
\bauthor{\bsnm{Brown},~\bfnm{Lawrence~D.}\binits{L.~D.}}
(\byear{1986}).
\btitle{Fundamentals of statistical exponential families with applications in statistical decision theory}.
\bseries{Institute of Mathematical Statistics Lecture Notes---Monograph Series}
\bvolume{9}.
\bpublisher{Institute of Mathematical Statistics, Hayward, CA}.
\bmrnumber{882001}
\end{bbook}
\endbibitem

\bibitem[\protect\citeauthoryear{Casella and Berger}{1990}]{casella:book}
\begin{bbook}[author]
\bauthor{\bsnm{Casella},~\bfnm{George}\binits{G.}} \AND \bauthor{\bsnm{Berger},~\bfnm{Roger~L.}\binits{R.~L.}}
(\byear{1990}).
\btitle{Statistical inference}.
\bseries{The Wadsworth \& Brooks/Cole Statistics/Probability Series}.
\bpublisher{Wadsworth \& Brooks/Cole Advanced Books \& Software, Pacific Grove, CA}.
\bmrnumber{1051420}
\end{bbook}
\endbibitem

\bibitem[\protect\citeauthoryear{Chen, Drton and Wang}{2019}]{Chen:19}
\begin{barticle}[author]
\bauthor{\bsnm{Chen},~\bfnm{Wenyu}\binits{W.}}, \bauthor{\bsnm{Drton},~\bfnm{Mathias}\binits{M.}} \AND \bauthor{\bsnm{Wang},~\bfnm{Y.~Samuel}\binits{Y.~S.}}
(\byear{2019}).
\btitle{On causal discovery with an equal-variance assumption}.
\bjournal{Biometrika}
\bvolume{106}
\bpages{973--980}.
\bmrnumber{4031210}
\end{barticle}
\endbibitem

\bibitem[\protect\citeauthoryear{Cox, Little and O'Shea}{2015}]{Cox:15}
\begin{bbook}[author]
\bauthor{\bsnm{Cox},~\bfnm{David~A.}\binits{D.~A.}}, \bauthor{\bsnm{Little},~\bfnm{John}\binits{J.}} \AND \bauthor{\bsnm{O'Shea},~\bfnm{Donal}\binits{D.}}
(\byear{2015}).
\btitle{Ideals, varieties, and algorithms},
\bedition{fourth} ed.
\bseries{Undergraduate Texts in Mathematics}.
\bpublisher{Springer, Cham}.
\bmrnumber{3330490}
\end{bbook}
\endbibitem

\bibitem[\protect\citeauthoryear{Drton}{2009}]{Drton:2009}
\begin{barticle}[author]
\bauthor{\bsnm{Drton},~\bfnm{Mathias}\binits{M.}}
(\byear{2009}).
\btitle{Likelihood ratio tests and singularities}.
\bjournal{Ann. Statist.}
\bvolume{37}
\bpages{979--1012}.
\bmrnumber{2502658}
\end{barticle}
\endbibitem

\bibitem[\protect\citeauthoryear{Drton}{2018}]{Drton:18}
\begin{bincollection}[author]
\bauthor{\bsnm{Drton},~\bfnm{Mathias}\binits{M.}}
(\byear{2018}).
\btitle{Algebraic problems in structural equation modeling}.
In \bbooktitle{The 50th anniversary of {G}r\"{o}bner bases}.
\bseries{Adv. Stud. Pure Math.}
\bvolume{77}
\bpages{35--86}.
\bpublisher{Math. Soc. Japan, Tokyo}.
\bmrnumber{3839705}
\end{bincollection}
\endbibitem

\bibitem[\protect\citeauthoryear{Drton and Maathuis}{2017}]{Drton:17}
\begin{barticle}[author]
\bauthor{\bsnm{Drton},~\bfnm{Mathias}\binits{M.}} \AND \bauthor{\bsnm{Maathuis},~\bfnm{Marloes~H.}\binits{M.~H.}}
(\byear{2017}).
\btitle{Structure Learning in Graphical Modeling}.
\bjournal{Annual Review of Statistics and Its Application}
\bvolume{4}
\bpages{365-393}.
\end{barticle}
\endbibitem

\bibitem[\protect\citeauthoryear{Ghoshal and Honorio}{2018}]{Ghoshal:18}
\begin{binproceedings}[author]
\bauthor{\bsnm{Ghoshal},~\bfnm{Asish}\binits{A.}} \AND \bauthor{\bsnm{Honorio},~\bfnm{Jean}\binits{J.}}
(\byear{2018}).
\btitle{Learning linear structural equation models in polynomial time and sample complexity}.
In \bbooktitle{International Conference on Artificial Intelligence and Statistics, {AISTATS}}.
\bseries{Proceedings of Machine Learning Research}
\bvolume{84}
\bpages{1466--1475}.
\bpublisher{{PMLR}}.
\end{binproceedings}
\endbibitem

\bibitem[\protect\citeauthoryear{Henckel, Perkovi\'{c} and Maathuis}{2022}]{Henckel:22}
\begin{barticle}[author]
\bauthor{\bsnm{Henckel},~\bfnm{Leonard}\binits{L.}}, \bauthor{\bsnm{Perkovi\'{c}},~\bfnm{Emilija}\binits{E.}} \AND \bauthor{\bsnm{Maathuis},~\bfnm{Marloes~H.}\binits{M.~H.}}
(\byear{2022}).
\btitle{Graphical criteria for efficient total effect estimation via adjustment in causal linear models}.
\bjournal{J. R. Stat. Soc. Ser. B. Stat. Methodol.}
\bvolume{84}
\bpages{579--599}.
\bmrnumber{4412998}
\end{barticle}
\endbibitem

\bibitem[\protect\citeauthoryear{Kauermann}{1996}]{Kauermann:96}
\begin{barticle}[author]
\bauthor{\bsnm{Kauermann},~\bfnm{G\"{o}ran}\binits{G.}}
(\byear{1996}).
\btitle{On a dualization of graphical {G}aussian models}.
\bjournal{Scand. J. Statist.}
\bvolume{23}
\bpages{105--116}.
\bmrnumber{1380485}
\end{barticle}
\endbibitem

\bibitem[\protect\citeauthoryear{Lauritzen}{1996}]{Lauritzen:book}
\begin{bbook}[author]
\bauthor{\bsnm{Lauritzen},~\bfnm{Steffen~L.}\binits{S.~L.}}
(\byear{1996}).
\btitle{Graphical models}.
\bseries{Oxford Statistical Science Series}
\bvolume{17}.
\bpublisher{The Clarendon Press, Oxford University Press, New York}.
\bmrnumber{1419991}
\end{bbook}
\endbibitem

\bibitem[\protect\citeauthoryear{Maathuis, Kalisch and B\"{u}hlmann}{2009}]{Maathuis:09}
\begin{barticle}[author]
\bauthor{\bsnm{Maathuis},~\bfnm{Marloes~H.}\binits{M.~H.}}, \bauthor{\bsnm{Kalisch},~\bfnm{Markus}\binits{M.}} \AND \bauthor{\bsnm{B\"{u}hlmann},~\bfnm{Peter}\binits{P.}}
(\byear{2009}).
\btitle{Estimating high-dimensional intervention effects from observational data}.
\bjournal{Ann. Statist.}
\bvolume{37}
\bpages{3133--3164}.
\bmrnumber{2549555}
\end{barticle}
\endbibitem

\bibitem[\protect\citeauthoryear{Maathuis et~al.}{2019}]{Handbook}
\begin{bbook}[author]
\beditor{\bsnm{Maathuis},~\bfnm{Marloes}\binits{M.}}, \beditor{\bsnm{Drton},~\bfnm{Mathias}\binits{M.}}, \beditor{\bsnm{Lauritzen},~\bfnm{Steffen}\binits{S.}} \AND \beditor{\bsnm{Wainwright},~\bfnm{Martin}\binits{M.}}, eds.
(\byear{2019}).
\btitle{Handbook of graphical models}.
\bseries{Chapman \& Hall/CRC Handbooks of Modern Statistical Methods}.
\bpublisher{CRC Press, Boca Raton, FL}.
\bmrnumber{3889064}
\end{bbook}
\endbibitem

\bibitem[\protect\citeauthoryear{Pearl}{2009}]{Pearl:book}
\begin{bbook}[author]
\bauthor{\bsnm{Pearl},~\bfnm{Judea}\binits{J.}}
(\byear{2009}).
\btitle{Causality},
\bedition{Second} ed.
\bpublisher{Cambridge University Press, Cambridge}.
\bmrnumber{2548166}
\end{bbook}
\endbibitem

\bibitem[\protect\citeauthoryear{Peters and B\"{u}hlmann}{2014}]{Peters:14}
\begin{barticle}[author]
\bauthor{\bsnm{Peters},~\bfnm{Jonas}\binits{J.}} \AND \bauthor{\bsnm{B\"{u}hlmann},~\bfnm{Peter}\binits{P.}}
(\byear{2014}).
\btitle{Identifiability of {G}aussian structural equation models with equal error variances}.
\bjournal{Biometrika}
\bvolume{101}
\bpages{219--228}.
\bmrnumber{3180667}
\end{barticle}
\endbibitem

\bibitem[\protect\citeauthoryear{Peters, Janzing and Sch\"{o}lkopf}{2017}]{Peters:book}
\begin{bbook}[author]
\bauthor{\bsnm{Peters},~\bfnm{Jonas}\binits{J.}}, \bauthor{\bsnm{Janzing},~\bfnm{Dominik}\binits{D.}} \AND \bauthor{\bsnm{Sch\"{o}lkopf},~\bfnm{Bernhard}\binits{B.}}
(\byear{2017}).
\btitle{Elements of causal inference}.
\bseries{Adaptive Computation and Machine Learning}.
\bpublisher{MIT Press, Cambridge, MA}.
\bmrnumber{3822088}
\end{bbook}
\endbibitem

\bibitem[\protect\citeauthoryear{Silander and Myllym{\"{a}}ki}{2006}]{Silander:06}
\begin{binproceedings}[author]
\bauthor{\bsnm{Silander},~\bfnm{Tomi}\binits{T.}} \AND \bauthor{\bsnm{Myllym{\"{a}}ki},~\bfnm{Petri}\binits{P.}}
(\byear{2006}).
\btitle{A Simple Approach for Finding the Globally Optimal Bayesian Network Structure}.
In \bbooktitle{Proceedings of the 22nd Conference in Uncertainty in Artificial Intelligence, {UAI}}.
\bpublisher{{AUAI} Press}.
\end{binproceedings}
\endbibitem

\bibitem[\protect\citeauthoryear{Spirtes, Glymour and Scheines}{2000}]{Spirtes:book}
\begin{bbook}[author]
\bauthor{\bsnm{Spirtes},~\bfnm{Peter}\binits{P.}}, \bauthor{\bsnm{Glymour},~\bfnm{Clark}\binits{C.}} \AND \bauthor{\bsnm{Scheines},~\bfnm{Richard}\binits{R.}}
(\byear{2000}).
\btitle{Causation, prediction, and search}.
\bseries{Adaptive Computation and Machine Learning}.
\bpublisher{MIT Press, Cambridge, MA}.
\bmrnumber{1815675}
\end{bbook}
\endbibitem

\bibitem[\protect\citeauthoryear{Strieder and Drton}{2023}]{Strieder:23}
\begin{barticle}[author]
\bauthor{\bsnm{Strieder},~\bfnm{David}\binits{D.}} \AND \bauthor{\bsnm{Drton},~\bfnm{Mathias}\binits{M.}}
(\byear{2023}).
\btitle{Confidence in causal inference under structure uncertainty in linear causal models with equal variances}.
\bjournal{J. Causal Inference}
\bvolume{11}.
\bmrnumber{4679813}
\end{barticle}
\endbibitem

\bibitem[\protect\citeauthoryear{Strieder and Drton}{2024}]{Strieder:24}
\begin{binproceedings}[author]
\bauthor{\bsnm{Strieder},~\bfnm{David}\binits{D.}} \AND \bauthor{\bsnm{Drton},~\bfnm{Mathias}\binits{M.}}
(\byear{2024}).
\btitle{Dual Likelihood for Causal Inference under Structure Uncertainty}.
In \bbooktitle{Causal Learning and Reasoning, {CLeaR}}.
\bseries{Proceedings of Machine Learning Research}
\bvolume{236}
\bpages{1--17}.
\bpublisher{{PMLR}}.
\end{binproceedings}
\endbibitem

\bibitem[\protect\citeauthoryear{Strieder et~al.}{2021}]{Strieder:21}
\begin{binproceedings}[author]
\bauthor{\bsnm{Strieder},~\bfnm{David}\binits{D.}}, \bauthor{\bsnm{Freidling},~\bfnm{Tobias}\binits{T.}}, \bauthor{\bsnm{Haffner},~\bfnm{Stefan}\binits{S.}} \AND \bauthor{\bsnm{Drton},~\bfnm{Mathias}\binits{M.}}
(\byear{2021}).
\btitle{Confidence in causal discovery with linear causal models}.
In \bbooktitle{Proceedings of the 37th Conference on Uncertainty in Artificial Intelligence, {UAI}}.
\bseries{Proceedings of Machine Learning Research}
\bvolume{161}
\bpages{1217--1226}.
\bpublisher{{AUAI} Press}.
\end{binproceedings}
\endbibitem

\bibitem[\protect\citeauthoryear{Wang, Kolar and Drton}{2023}]{Wang:23}
\begin{barticle}[author]
\bauthor{\bsnm{Wang},~\bfnm{Y.~Samuel}\binits{Y.~S.}}, \bauthor{\bsnm{Kolar},~\bfnm{Mladen}\binits{M.}} \AND \bauthor{\bsnm{Drton},~\bfnm{Mathias}\binits{M.}}
(\byear{2023}).
\btitle{Confidence Sets for Causal Orderings}.
\bjournal{arXiv preprint arXiv:2305.14506}.
\end{barticle}
\endbibitem

\bibitem[\protect\citeauthoryear{Wu and Drton}{2023}]{Wu:23}
\begin{barticle}[author]
\bauthor{\bsnm{Wu},~\bfnm{Jun}\binits{J.}} \AND \bauthor{\bsnm{Drton},~\bfnm{Mathias}\binits{M.}}
(\byear{2023}).
\btitle{Partial Homoscedasticity in Causal Discovery With Linear Models}.
\bjournal{{IEEE} J. Sel. Areas Inf. Theory}
\bvolume{4}
\bpages{639--650}.
\end{barticle}
\endbibitem

\end{thebibliography}

\newpage

\appendix

\section{Identifiable Example}\label{apd:example2}
Under the assumption of an underlying linear Gaussian SCM with partial homoscedasticity $\omega_1=\omega_2$, that is, $\Sigma_{1,1|p(1)}=\Sigma_{2,2|p(2)}$, the observational distribution $P_{\Sigma}$, given by 
 \begin{equation*}
      \Sigma=\begin{pmatrix} 1 & 1 & 2 \\ 
 1 & 2 & 3 \\
 2 & 3 & 5.5 \end{pmatrix},
  \end{equation*}
uniquely entails the total causal effect $\mathcal{C}(1 \rightarrow 2)=1$. This follows from the observation that $\Sigma_{1,1|p(1)}=\Sigma_{2,2|p(2)}$ if and only if $p(1)$ equals the empty set and $p(2)=\{1\}$. Thus, $\mathcal{C}(1 \rightarrow 2)=\Sigma_{1,2}(\Sigma_{1,1})^{-1}=1$.

\section{Non-identifiable Example}\label{apd:example}
In Example \ref{example}, we used computer algebra software to find a covariance matrix that satisfies both equal error variance constraints $\Sigma_{1,1}=\Sigma_{2,2|1}$ and $\Sigma_{1,1|\{2,3\}}=\Sigma_{2,2|3}$. The exact entries of the found covariance matrix are algebraic numbers, given as complex root expressions. Thus, we display in the following the found (exact) example up to a precision of $15$ decimal digits, that is, the observational distribution $P_{\Sigma}$, given by
  \begin{equation*}
      \Sigma=\begin{pmatrix} 1.00000000000000 & 0.294584930358565 & -0.176750958215139 \\ 
 0.294584930358565 & 1.08678028119436 & 0.648086846788844 \\
 -0.176750958215139 & 0.648086846788844 & 1.52145794696742 \end{pmatrix},
  \end{equation*}
could have been generate by two different linear Gaussian SCMs $\Vec{X}=B\Vec{X}+\Vec{\varepsilon}$, that is, either
  \begin{equation*}
      B_1=\begin{pmatrix} 0 & 0 & 0 \\ 
 0.294584930358565 & 0 & 0 \\
-0.383006074698015 & 0.700155015505460 & 0 \end{pmatrix}
  \end{equation*}
and $\text{Var}(\Vec{\varepsilon_1})=\text{diag}(1,1,1)$, represented by $G_1$ in Example \ref{example}, or 
  \begin{equation*}
      B_2=\begin{pmatrix} 0 & 
0.456230601077430 & -0.310510067542541 \\ 
 0 & 0 & 0.425964350891600 \\
0 & 0 & 0 \end{pmatrix}
  \end{equation*}
and $\text{Var}(\Vec{\varepsilon_2})=\text{diag}(0.810718388180567,0.810718388180567, 1.52145794696742)$, given by $G_2$.

\section{Proof of Theorem \ref{thm:confgen}}\label{apd:proofconfgen}

\begin{proof}
  Let $\psi \in \R$ and $\Sigma \in \mathcal{M}^{\psi}$. Since $\mathcal{M}^{\psi}=\bigcup_{G \in \mathcal{G}(d)} \mathcal{M}^{\psi}(G)$ there exists a complete graph $G$ such that  $\Sigma \in \mathcal{M}^{\psi}(G)$. Without loss of generality, we assume $ i <_G j$; otherwise, $\psi$ has to be zero, which is always contained in the proposed confidence regions by construction. 
  
  The model space $\mathcal{M}$ equals the set of all positive definite matrices PD$(d)$ and can be identified with an open subspace in $\R^{\frac{1}{2}(d^2+d)}$. Each (single) hypothesis of the union $\Sigma \in \mathcal{M}^{\psi}(G)$ then defines a $\frac{1}{2}(d^2+d)-1$-dimensional submanifold of $\R^{\frac{1}{2}(d^2+d)}$. Therefore, the asymptotic distribution of the dual likelihood ratio test statistic under every single hypothesis is given by $ \chi^2_1$ and, thus, 
  \begin{equation*}
        P_{\Sigma}\Bigg( n\bigg(\sup_{\Sigma \in \mathcal{M}} \ell^{dual}_n(\Sigma)- \sup_{\Sigma \in  \mathcal{M}^{\psi}(G)} \ell^{dual}_n(\Sigma)\bigg)> \chi_{1,1-\alpha}^2 \Bigg) \rightarrow \alpha.
  \end{equation*}
  
  For details on this classical result about the limit distribution of likelihood ratio test for submanifolds and the transformation to dual likelihood, we refer to \citet{Drton:2009} and \citet{Strieder:24}. Furthermore, dual likelihood estimation via solving a sequence of linear regression problems (see Section \ref{sec:estimation}) yields the following optimal values
  \begin{align*}
      \sup_{\Sigma \in \mathcal{M}} \ell^{dual}_n(\Sigma)=&  \sup_{\Sigma \in \mathcal{M}(G)} \ell^{dual}_n(\Sigma)=-\sum_{k=1}^d \log\big((\widehat{\Sigma}^{-1})_{k,k|d(k)}\big)-d, \\
      \sup_{\Sigma \in  \mathcal{M}^{\psi}(G)}\ell^{dual}_n(\Sigma) =  &-   \sum_{k\neq i}^{d}\log\big((\widehat{\Sigma}^{-1})_{k,k|d(k)}\big) \\ &-\log\big((\widehat{\Sigma}^{-1})_{i,i|d(i)\setminus \{j\}}+ \psi^2 (\widehat{\Sigma}^{-1})_{j,j|d(i)\setminus \{j\}}  + 2\psi (\widehat{\Sigma}^{-1})_{i,j|d(i)\setminus \{j\}}\big) - d . 
  \end{align*}      
  Plugging in these dual likelihood estimates, we view
    \begin{equation*}
      n\bigg(\sup_{\Sigma \in \mathcal{M}} \ell^{dual}_n(\Sigma)- \sup_{\Sigma \in  \mathcal{M}^{\psi}(G)} \ell^{dual}_n(\Sigma)\bigg)-\chi_{1,1-\alpha}^2
    \end{equation*}
     as a strictly convex quadratic polynomial in $\psi$, which has real roots if $D(G) \geq 0$, where 
    \begin{equation*}
        D(G):= (\widehat{\Sigma}^{-1})_{j,i|d(i)\setminus \{j\}}^2-(\widehat{\Sigma}^{-1})_{j,j|d(i)\setminus \{j\}}\Big((\widehat{\Sigma}^{-1})_{i,i|d(i)\setminus \{j\}} - (\widehat{\Sigma}^{-1})_{i,i|d(i)} \exp\big(\tfrac{\chi^2_{1,1-\alpha}}{n}\big) \Big).
    \end{equation*}
    Thus, the inequality $ n\big(\sup_{\Sigma \in \mathcal{M}} \ell^{dual}_n(\Sigma)- \sup_{\Sigma \in  \mathcal{M}^{\psi}(G)} \ell^{dual}_n(\Sigma)\big)> \chi_{1,1-\alpha}^2$ holds if and only if
    \begin{equation*} 
        \psi \in \Bigg[\frac{-(\widehat{\Sigma}^{-1})_{i,j|d(i)\setminus \{j\}}  - \sqrt{D(G)}}{(\widehat{\Sigma}^{-1})_{j,j|d(i)\setminus \{j\}}}, \frac{-(\widehat{\Sigma}^{-1})_{i,j|d(i)\setminus \{j\}}  + \sqrt{D(G)}}{(\widehat{\Sigma}^{-1})_{j,j|d(i)\setminus \{j\}}}  \Bigg].
    \end{equation*}
 The test inversion approach and intersection union theory complete the proof.
 \end{proof}

\section{Proof of Theorem \ref{thm:confpart}}\label{apd:proofconfpart}

\begin{proof}
  Let $\psi \in \R$ and $\Sigma \in \mathcal{M}_{PEV}^{\psi}$. Since $\mathcal{M}_{PEV}^{\psi}=\bigcup_{G \in \mathcal{G}(d)} \mathcal{M}_{PEV}^{\psi}(G)$ there exists a complete graph $G$, such that  $\Sigma \in \mathcal{M}_{PEV}^{\psi}(G)$. We assume $ i <_G j$; the case $j <_G i$ can be constructed analogously. 

  The dual likelihood ratio test statistic is upper bounded by the larger test statistic of testing against the entire cone of positive definite matrices, that is,  
    \begin{equation*}
         n\bigg(\sup_{\Sigma \in \mathcal{M}_{PEV}} \ell^{dual}_n(\Sigma)- \sup_{\Sigma \in  \mathcal{M}_{PEV}^{\psi}(G)} \ell^{dual}_n(\Sigma)\bigg) \leq  n\bigg(\sup_{\Sigma \in \text{PD}(d)} \ell^{dual}_n(\Sigma)- \sup_{\Sigma \in  \mathcal{M}_{PEV}^{\psi}(G)} \ell^{dual}_n(\Sigma)\bigg).
    \end{equation*}
  
 Within the set of all positive definite matrices PD$(d)$, each (single) hypothesis of the union $\Sigma \in \mathcal{M}_{PEV}^{\psi}(G)$ defines a $\frac{1}{2}(d^2+d)-2$-dimensional submanifold of $\R^{\frac{1}{2}(d^2+d)}$. Therefore, the asymptotic distribution of this stochastic upper bound of the dual likelihood ratio test statistic is given by $ \chi^2_2$ and, thus, with the conservative critical value $\chi_{2,1-\alpha}^2$ from the stochastic upper bound, we obtain 
  \begin{align*}
          &P_{\Sigma}\Bigg( n\bigg(\sup_{\Sigma \in \mathcal{M}_{PEV}} \ell^{dual}_n(\Sigma)- \sup_{\Sigma \in  \mathcal{M}_{PEV}^{\psi}(G)} \ell^{dual}_n(\Sigma)\bigg)> \chi_{2,1-\alpha}^2 \Bigg) \\ &\leq P_{\Sigma}\Bigg( n\bigg(\sup_{\Sigma \in \text{PD}(d)} \ell^{dual}_n(\Sigma)- \sup_{\Sigma \in  \mathcal{M}_{PEV}^{\psi}(G)} \ell^{dual}_n(\Sigma)\bigg)> \chi_{2,1-\alpha}^2 \Bigg) \rightarrow \alpha.
  \end{align*}
  
  For details on the limit distribution of the likelihood ratio test for submanifolds and the transformation to dual likelihood, we refer to \citet{Drton:2009} and \citet{Strieder:24}. Furthermore, the dual likelihood estimation via solving a sequence of linear regression problems (see Section \ref{sec:estimation}) yields 

  \begin{equation*}
      \sup_{\Sigma \in \mathcal{M}_{PEV}} \ell^{dual}_n(\Sigma)=-2\log(\tfrac{K}{2}) -d. 
  \end{equation*}
  with 
 \begin{equation*}
        K:=\min_{G \in \mathcal{G}(d)} \prod^d_{k\neq i,j}\sqrt{(\widehat{\Sigma}^{-1})_{k,k|d(k)} } \Big((\widehat{\Sigma}^{-1})_{i,i|d(i)}+(\widehat{\Sigma}^{-1})_{j,j|d(j)}\Big),
    \end{equation*}
as well as 
\begin{align*}
\sup_{\Sigma \in \mathcal{M}_{PEV}^{\psi}(G)} \ell^{dual}_n(\Sigma)= &-\sum_{k \neq i,j}^d\log\big((\widehat{\Sigma}^{-1})_{k,k|d(k)}\big) -d -2\log\Big(\tfrac{1}{2}\big((\widehat{\Sigma}^{-1})_{i,i|d(i)\setminus \{j\}} \\ &+ \psi^2 (\widehat{\Sigma}^{-1})_{j,j|d(i)\setminus \{j\}}  + 2\psi (\widehat{\Sigma}^{-1})_{i,j|d(i)\setminus \{j\}}+(\widehat{\Sigma}^{-1})_{j,j|d(j)}\big)\Big).
\end{align*}
 Plugging in these dual likelihood estimates, we view
    \begin{equation*}
      n\bigg(\sup_{\Sigma \in \mathcal{M}_{PEV}} \ell^{dual}_n(\Sigma)- \sup_{\Sigma \in  \mathcal{M}_{PEV}^{\psi}(G)} \ell^{dual}_n(\Sigma)\bigg)- \chi_{2,1-\alpha}^2 
    \end{equation*}
     as a strictly convex quadratic polynomial in $\psi$, which has real roots if $D(G) \geq 0$, where 
    \begin{align*}
        D(G):= &(\widehat{\Sigma}^{-1})_{j,i|d(i)\setminus \{j\}}^2\\ &-(\widehat{\Sigma}^{-1})_{j,j|d(i)\setminus \{j\}}\Bigg((\widehat{\Sigma}^{-1})_{j,j|d(j)}+(\widehat{\Sigma}^{-1})_{i,i|d(i)\setminus \{j\}} -\frac{K \exp\big(\tfrac{\chi^2_{2,1-\alpha}}{2n}\big)}{\prod^d_{k\neq i,j}\sqrt{(\widehat{\Sigma}^{-1})_{k,k|d(k)} }} \Bigg)
    \end{align*}
    Thus, the inequality $ n\big(\sup_{\Sigma \in \mathcal{M}_{PEV}} \ell^{dual}_n(\Sigma)- \sup_{\Sigma \in  \mathcal{M}_{PEV}^{\psi}(G)} \ell^{dual}_n(\Sigma)\big)> \chi_{2,1-\alpha}^2 $ holds if and only if
    \begin{equation*} 
        \psi \in \Bigg[\frac{-(\widehat{\Sigma}^{-1})_{i,j|d(i)\setminus \{j\}}  - \sqrt{D(G)}}{(\widehat{\Sigma}^{-1})_{j,j|d(i)\setminus \{j\}}}, \frac{-(\widehat{\Sigma}^{-1})_{i,j|d(i)\setminus \{j\}}  + \sqrt{D(G)}}{(\widehat{\Sigma}^{-1})_{j,j|d(i)\setminus \{j\}}}  \Bigg].
    \end{equation*}
    The test inversion approach and intersection union theory complete the proof.
\end{proof}

\section{Additional Simulation Results}\label{apd:sim}

In this section, we recapitulate the synthetic data generation process adapted from \citet{Strieder:24} and present further simulation results for true non-zero effects. To generate a synthetic data set, we first select a random permutation of $10$ nodes. Second, we prune the corresponding complete graph by including each edge with a probability of $0.5$. Then, we generate edge weights according to a normal distribution $\mathcal{N}(0.5,0.1)$ and generate $n$ samples according to the linear Gaussian SCM represented by the selected DAG. Here, we use three different error variance regimes, that is, \texttt{general data} where each error variance is sampled uniformly from $[0.5,1.5]$, \texttt{partial ev data} with the two error variance of potential causal and effect being $1$ and the rest sampled uniformly from $[0.5,1.5]$, and \texttt{ev data} with all error variances being $1$. We repeated this process to obtain $1000$ independent data sets (twice, with true non-zero effect and no effect) to calculate the presented empirical quantities. 

\begin{figure}[t]
    \centering
    \includegraphics[width=0.9\linewidth]{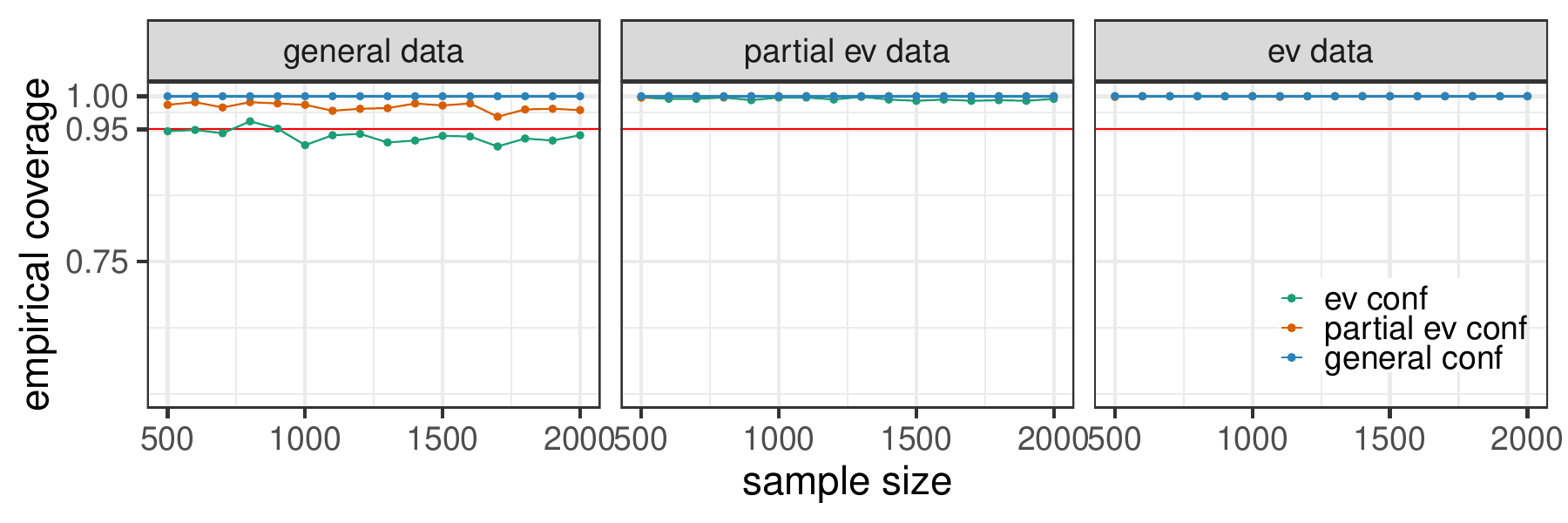}
    \caption{Empirical coverage of $95\%$-confidence intervals for the total causal effect.}\label{fig:coverfalse10}
\end{figure}

\begin{figure}[t]
    \centering
    \includegraphics[width=0.9\linewidth]{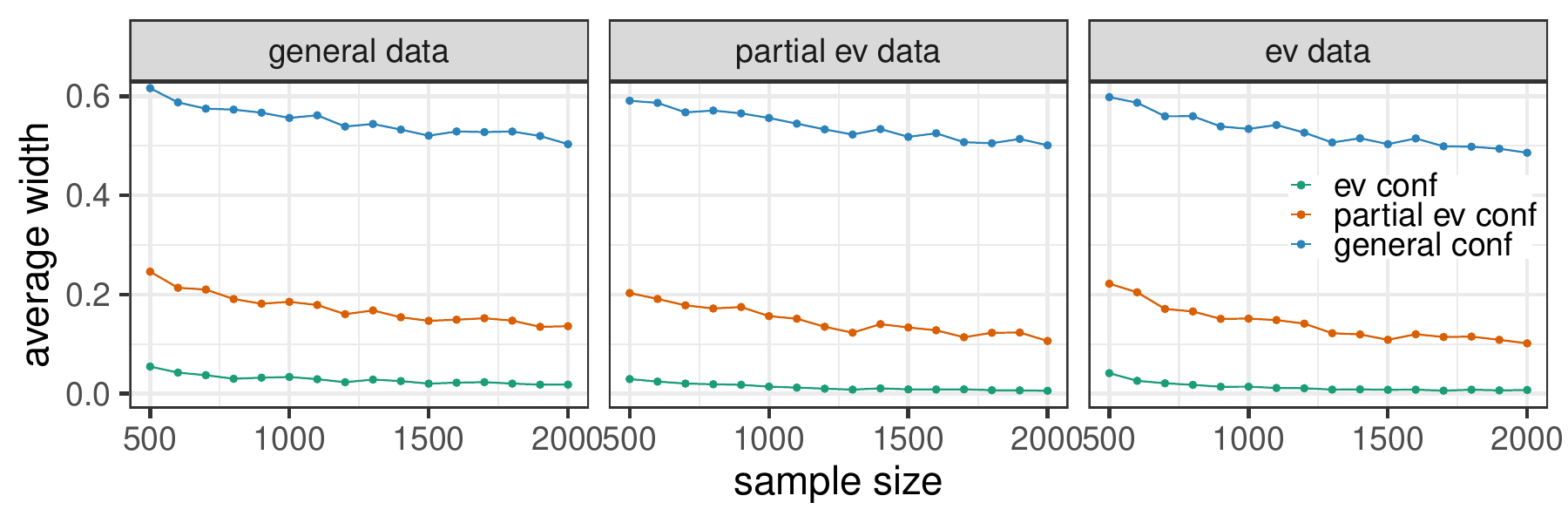}
    \caption{Mean width of $95\%$-confidence intervals for the total causal effect.}\label{fig:widthfalse10}
\end{figure}

Figure \ref{fig:coverfalse10} shows the empirical coverage frequencies when no true effect exists. Note that this task is 'easier' as no effect corresponds to the larger class of models being plausible where node $j$ proceeds node $i$ in the causal ordering. However, the information contained in the intervals still significantly differs, highlighted by the average width reported in Figure \ref{fig:widthfalse10}. Finally, note that, in this setting, the zero proportions are equal to the empirical coverage.


\end{document}